\documentclass[aps,superscriptaddress,twocolumn]{revtex4}
\usepackage{amsmath}
\usepackage{graphicx}
\usepackage{dcolumn}
\usepackage{verbatim}
\usepackage{times}
\usepackage{subfigure}
\usepackage{bm}
\usepackage{color}
\usepackage[colorlinks,citecolor=blue,linkcolor=blue,dvipdfm]{hyperref}
\usepackage{subfigure}

\usepackage{booktabs}
\usepackage{appendix}

\begin{document}

\title{Correlated Dirac semimetal states in nonsymmorphic MIrO$_3$ (M=Sr, Ba and Ca)}

\author{Zhi-Ming Yang}
\affiliation{College of Physics and Electronic Information Engineering, Guilin University of Technology, Guilin 541004, China}
\author{Huan Li}
\email{lihuan@glut.edu.cn}
\affiliation{College of Physics and Electronic Information Engineering, Guilin University of Technology, Guilin 541004, China}

\date{\today}

\begin{abstract}

Nonsymmorphic symmetries can give rise to Dirac semimetal (DSM) states. However, few studies have been conducted on DSMs in interacting systems. Here, we induce interacting DSM states in nonsymmorphic iridium oxides SrIrO$_3$, BaIrO$_3$ and CaIrO$_3$, and contend that the interaction of electron-electron correlations, strong spin-orbital coupling, and symmetry protection can drive robust and exotic DSM states. Based on the density functional theory combined with dynamical mean-field theory (DFT + DMFT), with the Coulomb interaction parameters computed through doubly screened Coulomb correction approach, we discover that the Dirac fermions are constituted by the strongly spin-orbital coupled $J_{\mathrm{eff}} = 1/2$ states resulting from $t_{2g}$ orbits of Ir, with significant mass enhancement. Moreover, the nonsymmorphic symmetries induce topological surface bands and Fermi arcs on the (001) surface, which are well separated from bulk states. Our findings establish nonsymmorphic iridium oxides as correlated DSMs under strong electron-electron and spin-orbital interactions.

\end{abstract}

\maketitle

\section{Introduction}

Topological semimetals, encompassing Dirac semimetals and Weyl semimetals, constitute a class of topological states that have garnered widespread attention~\cite{Wang12,Wang13,Lv15,Wang19}. For these topological semimetals, the involvement of electronic correlations frequently has a significant impact on the behavior of Dirac or Weyl quasi-particles, giving rise to novel transport behaviors such as giant spontaneous Hall effect, thus making the search for topological nodes rather arduous~\cite{Lai18,Cao20,Kofuji21,Dzsaber21}. Consequently, there are scarce reports of interacting Dirac semimetals in real materials~\cite{Ma23}, and the search for Dirac semimetals in correlated materials holds both theoretical and experimental significance.

In $5d$ transition metal compounds, the intensities of electronic correlation, crystal field splitting, and spin-orbit coupling (SOC) are comparable. Under the combined influence of these interactions, the system demonstrates various distinctive behaviors. For instance, in iridium salts, the Ir-$5d$ electrons have a strong SOC, which leads to the recombination and splitting of the $t_{2g}$ energy level in the oxygen octahedron into two orbitals with effective angular momenta of $J_{\mathrm{eff}}$=1/2 and 3/2~\cite{Kim08}. Here, the $J_{\mathrm{eff}}$=1/2 orbital is half-filled and is readily influenced by electron correlation. The calculation based on the dynamic mean-field theory (DMFT) indicates that in numerous iridium salts, the $J_{\mathrm{eff}}$=1/2 orbital is inclined to undergo Mott transition and even magnetic phase transition~\cite{Cao18,Samanta24,Fujiyama12,Zhang13}. Thus, $5d$ transition metal compounds can act as ideal research platform for various competitive interactions.

Theoretical and experimental studies have revealed that nonsymmorphic symmetry can generate band overlap at certain high symmetry points in the Brillouin zone~\cite{Young15}. With the introduction of SOC, these quadruple-degenerate points can support Dirac nodes~\cite{Fu17,Schoop15,Chen17,Wu22}. For some nonsymmorphic materials of iridium oxides, due to their strong SOC, they can also hold Dirac semimetal states~\cite{Takayama18}. Moreover, the presence of strong electron correlation in iridium oxides can hold Dirac quasiparticles with strong correlation character~\cite{Hsu21}.
In this context, we ascertained that three iridium oxides SrIrO$_3$, BaIrO$_3$ and CaIrO$_3$ possessing nonsymmorphic symmetries can serve as ideal platforms for the study of correlated Dirac semimetals. We employed the DFT + DMFT theory to calculate the electronic correlations and discovered that a strong SOC was capable of reorganizing the electron orbitals near the Fermi level to generate $J_{\mathrm{eff}} = 1/2$ orbital and produce Dirac nodes at certain high symmetry points in the Brillouin zone under the protection of nonsymmorphic symmetries. Furthermore, these Dirac quasi-particles display a significant mass enhancement effect, reflecting their interacting intrinsicality. We also found that the surface states of SrIrO$_3$ and BaIrO$_3$ are well separated from the bulk states, featuring robust surface-state Fermi arcs that can be observed in future experiments.

\begin{figure}[tbp]
\hspace{-0cm} \includegraphics[totalheight=1.5in]{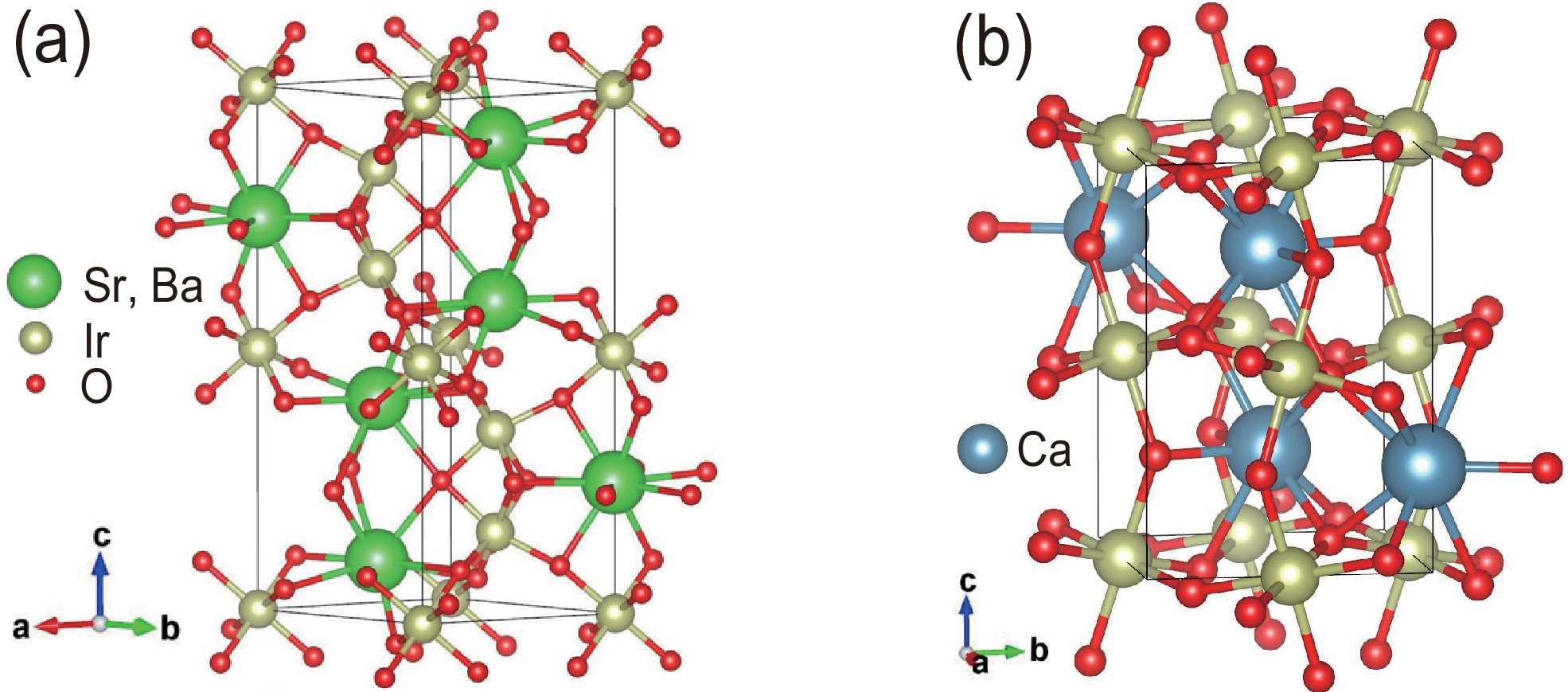}
\caption{Crystal structures of iridium oxides (a) SrIrO$_3$, BaIrO$_3$, and (b) CaIrO$_3$.
}
\label{lattice}
\end{figure}

\section{crystal structure and computational methods}

\begin{figure*}[tbp]
\hspace{-0.5cm} \includegraphics[totalheight=1.7in]{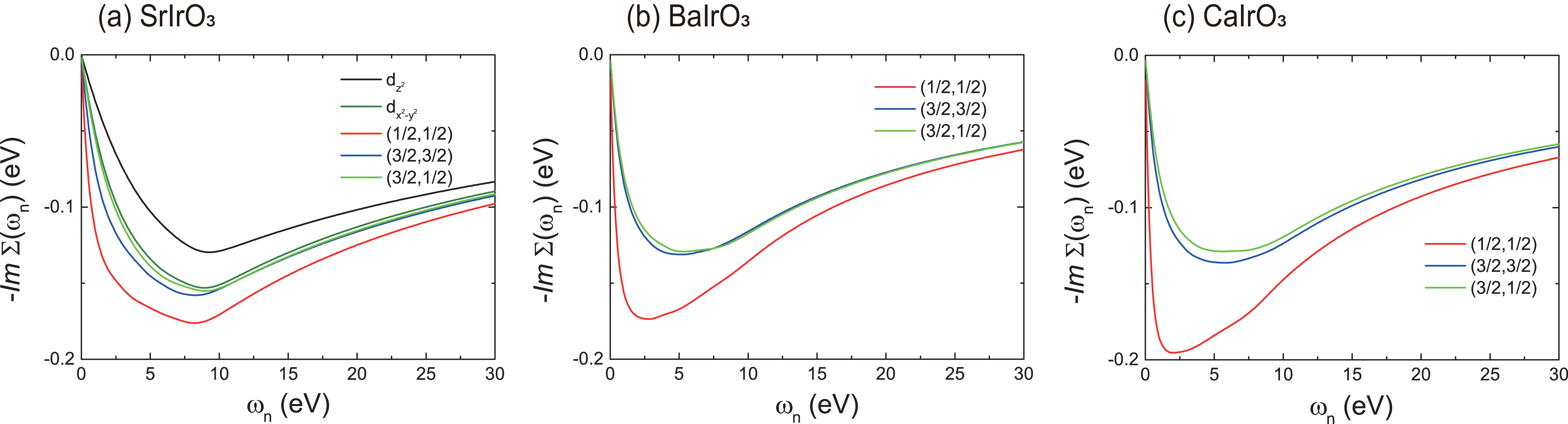}
\caption{DFT+DMFT self-energies of (a) SrIrO$_3$ and (b) BaIrO$_3$ at 116 K. The self-energies of different orbitals are displayed with lines of different colors.
}
\label{sig}
\end{figure*}

Fig.\ref{lattice} depicts the crystal structures of SrIrO$_3$, BaIrO$_3$ and CaIrO$_3$. SrIrO$_3$ and BaIrO$_3$ pertain to the monoclinic crystal structure with $C2/c$ space group (no. 15)~\cite{Hsu21} and can be regarded as being distorted from the hexagonal perovskite structure. Evidently, this lattice structure possesses nonsymmorphic symmetries, namely gliding mirror planes and screw axes. Specifically, the glide plane lies on the plane expanded by the vectors $\mathbf{a}+\mathbf{b}$ and $\mathbf{c}$, and the glide vector is (0, 0, 1/2). CaIrO$_3$ crystallizes in an orthorhombic perovskite structure, with $pnma$ space group (no. 62)~\cite{Tsuchiya07}, and possesses multiple glide planes and screw axes~\cite{Zeb12}.
The $5d$ transition metal compounds iridates possess a strong SOC in Ir atoms, under the influence of SOC, the nonsymmorphic lattice symmetries can permit the emergence of Dirac nodes at some high-symmetric points in the Brillouin zone, as will be discussed in the following text.

To further explore the impact of electronic correlation on iridium oxides, our study adopted a combined approach of density functional theory (DFT) and dynamical mean-field theory (DMFT), and employed the EDMFT software package for such computations~\cite{Haule10}. In the DFT section, we utilized the WIEN2k code to generate the Kohn-Sham single-particle Hamiltonian $\mathbf{H}_{\mathrm{KS}}$ via the full-potential linear augmented plane-wave method~\cite{Blaha20}. Subsequently, this Hamiltonian was combined with the interaction term $\mathbf{H}_{\mathrm{int}}$ and the double-counting term $\sum_{\mathrm{dc}}$ to obtain the lattice model $\mathbf{H}_{\mathrm{DFT+DMFT}}$ = $\mathbf{H}_{\mathrm{KS}}$ + $\mathbf{H}_{\mathrm{int}}$ - $\sum_{\mathrm{dc}}$. The DMFT algorithm projected the states within an energy window [-10, 10] eV relative to the Fermi level in the lattice model onto a single-site Anderson impurity model and employed the hybridization expansion version of continuous-time quantum Monte Carlo (CT-QMC) method to solve this impurity model. To acquire the real-frequency self-energy of $5d$ electrons, the maximum-entropy method was utilized for analytical continuation, converting the complex-frequency results obtained through CT-QMC simulations into real frequencies. The iterative DFT+DMFT calculation approach enables the achievement of complete charge self-consistency, and this computational method has been extensively applied in the investigation of electronic correlations in transition metal compounds~\cite{Arita12,Zhang13,Cao18,Tian24}.

In the DFT portion, a 15$\times$15$\times$6$\mathbf{k}$ grid was employed for the integration of the Brillouin zone, and the cutoff parameter $K_{\mathrm{max}}$ was set according to $R_{\mathrm{MT}}K_{\mathrm{max}}$ = 7.0. Additionally, SOC was consistently incorporated into the calculations as it plays a crucial role in the electronic structure of iridates. To validate the accuracy and reliability of the DFT band structure, the VASP code was utilized for cross-validation. To better conform to experimental observations, specific values of local Coulomb repulsion energy $U$ and the Hund coupling $J$ were adopted via doubly screened Coulomb correction approach (DSCC)~\cite{Liu23}, which are in accordance with previous literature reports of related iridium oxides~\cite{Arita12,Zhang13,Zhang17}. Furthermore, we also utilized various values of $U$ and $J$ to examine the degree of dependence of electron correlation on the electronic structure. The double-counting term was implemented in a nominal manner, namely $\sum_{\mathrm{dc}}$ = $U(n_f - 1/2)$ $-$ $J/2(n_f - 1)$, where $n_f$ represents the average occupancy of the Ir-$5d$ orbitals, which is approximately $5$ in iridates. Each iteration of the DFT+DMFT process consists of one DMFT calculation followed by 20 DFT calculations. In each CT-QMC computation, a total of 128 CPU cores were utilized to execute a significant number of (10 $\times$ 10$^8$) QMC steps. Typically, after approximately 30 to 40 iterations of the combination of DFT and DMFT, the self-consistency condition is attained, followed by five additional iterations to further average the self-energies.

\heavyrulewidth=1bp

\begin{table*}
\small
\renewcommand\arraystretch{1.3}
\caption{\label{tab1}
Mass enhancement $\frac{m^*}{m_{\mathrm{DFT}}}$ of Dirac quasi-particles under DFT+DMFT simulations at 116 K, with $(U,J)_{\mathrm{DSCC}}$=(2.16, 0.61), (2.37, 0.64) and (2.42. 0.63) eV for SrIrO$_3$, BaIrO$_3$ and CaIrO$_3$, respectively. For comparison, larger values of $U$ and $J$ are also employed in DFT+DMFT calculation.}
\begin{tabular*}{17cm}{@{\extracolsep{\fill}}ccccccccc}
\toprule
 $(U,J)$       & use DSCC          &  use $(4.5,0.8)$ eV & Exp.  \\
\hline
  SrIrO$_3$ & 1.25  & 1.63& 1.4~\cite{Hsu21}  \\
  BaIrO$_3$ & 1.49 & 2.89& -  \\
  CaIrO$_3$ & 1.53  & 3.12  & -  \\

 \bottomrule
\end{tabular*}
\label{data}
\end{table*}

\heavyrulewidth=1bp

\begin{table}
\small
\renewcommand\arraystretch{1.3}
\caption{\label{tab2}
Temperature dependence of mass enhancement of Dirac quasi-particles using $(U,J)_{\mathrm{DSCC}}$ in DFT+DMFT simulations.}
\begin{tabular*}{8cm}{@{\extracolsep{\fill}}ccccccccc}
\toprule
 Temperature (K)       &   116 & 400 \\
\hline
 SrIrO$_3$ &  1.25& 1.24  \\
 BaIrO$_3$ &   1.49&  1.46 \\
 CaIrO$_3$ &   1.53&  1.55 \\
 \bottomrule
\end{tabular*}
\label{data}
\end{table}

\section{spin-orbit coupling and electronic correlations}

\begin{figure*}[tbp]
\hspace{0cm} \includegraphics[totalheight=2.8in]{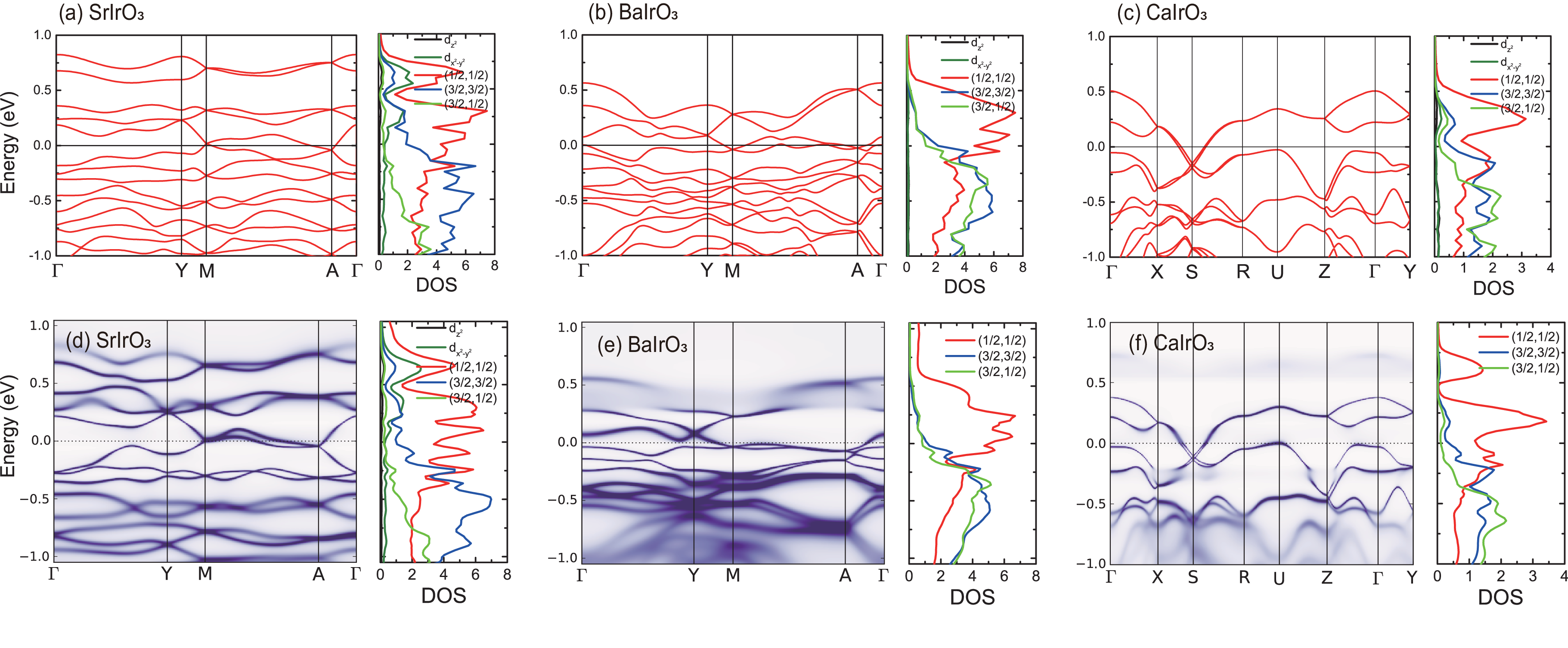}
\caption{ (a), (b) and (c) represent the DFT energy bands of SrIrO$_3$, BaIrO$_3$ and CaIrO$_3$ respectively, while (d) to (f) represent their DFT+DMFT spectral functions. The right panel in each graph shows the corresponding partial density of states projected into ($J_{\mathrm{eff}}, J_\mathrm{z}$) states.
}
\label{bands_specfuncs}
\end{figure*}

In iridates, under the crystal field effect of the IrO$_6$ octahedrons, the $d$ orbitals of Ir are split into $e_g$ and $t_{2g}$ orbitals. The $e_g$ orbitals are doubly degenerate with a relatively high energy, while the $t_{2g}$ orbitals are triply degenerate and their energy levels are close to the Fermi energy. However, since the strength of SOC in iridates is comparable to that of the crystal field, the $t_{2g}$ orbitals will be reorganized into three orbitals, with effective angular momenta of ($J_{\mathrm{eff}}, J_\mathrm{z}$) = (1/2, $\pm$1/2), (3/2, $\pm$3/2), and (3/2, $\pm$1/2), respectively~\cite{Kim08}. Near the Fermi level, the $J_{\mathrm{eff}}$ = 1/2 orbital dominates and can be significantly influenced by Coulomb correlations, resulting in the singular effects such as Mott transitions or magnetic phase transitions~\cite{Cao18,Samanta24,Fujiyama12,Zhang13,Imai22,Kaib22,Arita12,Zhang22,Nawa21,Kim15}.

We employed the Wien2k software to calculate the DFT electron bands of MIrO$_3$ (M=Sr, Ba, and Ca) and carried out bidirectional verification using the VASP package. It can be observed from Fig. \ref{bands_specfuncs}(a) to (c) that, due to nonsymmorphic symmetry, band crossings occur at the M and A points in SrIrO$_3$ and BaIrO$_3$, while for CaIrO$_3$, the band crossings take place near S point.
Given that the lattices possess spatial inversion symmetry, each band is doubly degenerate under the SOC effect. Hence, these crossing points are quadruply degenerate Dirac points~\cite{Young15}. To analyze the orbital components near the Dirac points, we diagonalized the matrix of impurity levels corresponding to the $d$ electrons, thereby obtaining the correct impurity orbitals near the Fermi level, namely $(J_{\mathrm{eff}}, J_\mathrm{z})$ states, and expressed them as a linear combination of spheric harmonics, enabling us to obtain the partial DOS of the $(J_{\mathrm{eff}}, J_\mathrm{z})$ orbitals~\cite{Haule10}. The results are presented in the smaller plots on the right panels in Fig. \ref{bands_specfuncs}(a) to (c). It can be discerned that due to the interplay between SOC and the crystal-field splitting, the DOS near the Fermi level is mainly composed by the $(J_{\mathrm{eff}}, J_\mathrm{z})=(1/2, 1/2)$ state, with a small amount of $(3/2, 3/2)$ component (approximately half of the $(1/2, 1/2)$ component), while the contributions from the remaining orbitals $((3/2, 1/2)$ and the two $e_g$ orbitals) are essentially negligible.
Compared with $J_{\mathrm{eff}}=3/2$ orbitals that are almost completely filled, the $J_{\mathrm{eff}}= 1/2$ orbital is essentially half-filled, thus the occupation of $5d$ electrons is $5$ (i.e., $t^5_{2g}$ configuration), corresponding to Ir$^{4+}$ valence. Therefore, $J_{\mathrm{eff}}= 1/2$ orbital is more susceptible to Coulomb correlation, and may induce the Mott transition and magnetic transition in some iridates~\cite{Cao18,Samanta24,Fujiyama12,Zhang13,Imai22,Kaib22,Arita12,Zhang22,Nawa21,Kim15}.

\begin{figure*}[tbp]
\hspace{-2cm} \includegraphics[totalheight=2.8in]{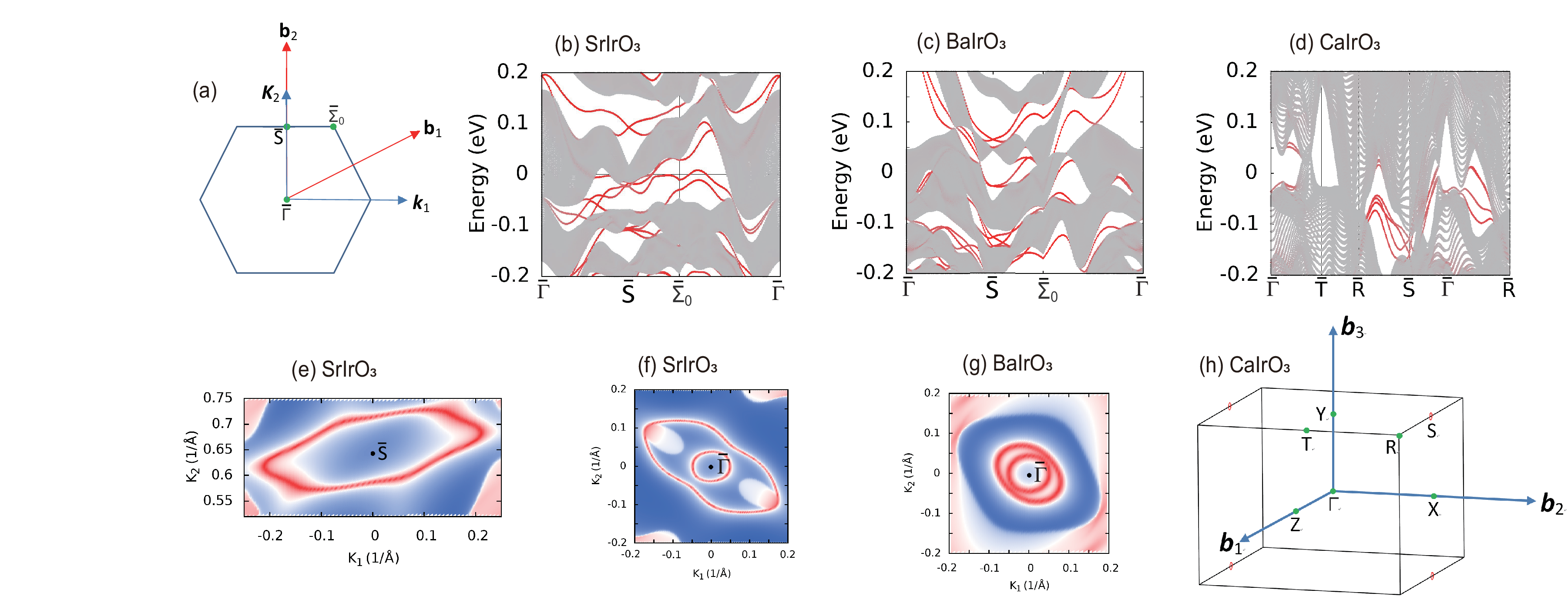}
\caption{(a) The (001) surface Brillouin zone of SrIrO$_3$ and BaIrO$_3$. (b) to (d) represent the slab energy spectra of SrIrO$_3$, BaIrO$_3$ and CaIrO$_3$ respectively, among which the red lines denote the surface states. (e) and (f) show the surface-state Fermi arcs of SrIrO$_3$ emerging near the $\bar{S}$ and $\bar{\Gamma}$ points, at a energy cut at 0.11 eV and 0.18 eV, respectively. (g) The surface-state Fermi arcs of BaIrO$_3$ emerging near the $\bar{\Gamma}$ point, at a energy cut at 0.254 eV. (h) Dirac node rings of CaIrO$_3$, which are denoted by small red circles near the Brillouin zone boundaries.
}
\label{Slab}
\end{figure*}

To investigate the consequence of $d$-electron Coulomb correlation on the Dirac quasiparticles in SrIrO$_3$, BaIrO$_3$ and CaIrO$_3$, we adopted the DFT + DMFT algorithm~\cite{Haule10} and calculated the self-energies of each $d$ orbitals.
The Coulomb interaction parameters, such as on-site Coulomb repulsion $U$ and Hund's coupling $J$, play a crucial role in the DFT + DMFT simulation. To calculate these parameters efficiently and precisely, we utilized the doubly screened Coulomb correction approach~\cite{Liu23}. The calculation results are ($U$, $J$) = (2.16, 0.61), (2.37, 0.64) and (2.42, 0.63) eV for SrIrO$_3$, BaIrO$_3$ and CaIrO$_3$. The larger lattice parameters of BaIrO$_3$ compared to SrIrO$_3$ enhances the localization of Ir-5$d$ electrons and consequently leads to an larger Hubbard parameter $U$. It can be noted that the obtained values are comparable to that adopted by the constrained random phase approximation
(cRPA) method for Sr$_2$IrO$_4$ and Ba$_2$IrO$_4$~\cite{Arita12,Zhang13,Zhang17}.
The correlation effect can be gauged by the mass enhancement factor $Z=\frac{m^*}{m_{\mathrm{DFT}}}=1-\frac{\partial \mathrm{Im} \Sigma(\omega_n)}{\partial\omega_n}|_{\omega_n\rightarrow0}$.
As can be seen from Fig. \ref{sig}, the $J_{\mathrm{eff}}= 1/2$ orbital in both iridates has a more pronounced mass enhancement than other orbitals. Moreover, the mass enhancement of $J_{\mathrm{eff}}= 1/2$ orbital in isomorphic BaIrO$_3$ is stronger than that of SrIrO$_3$ ($Z$=1.49 vs 1.25), demonstrating that the correlation effect of Dirac quasiparticles is more remarkable in BaIrO$_3$. In addition, we evaluated the mass enhancement factors across a range of temperatures and observed that their variation with temperature was relatively smooth, exhibiting only a minor increase upon cooling, see Tab. \ref{tab2}. Consequently, the mass enhancement result at 116 K is sufficiently representative of the characteristics under low-temperature conditions.

In some DFT+DMFT studies, larger correlation parameters $U$ were employed for the calculations of iridates and ruthenium trihalides~\cite{Samanta24}. To systematically investigate the impact of the choice of correlation parameters on DFT+DMFT calculations, we evaluated the mass enhancement factor under increased $U=4.5$ eV. The detailed results are presented in Tab. \ref{tab1}. It is evident that this leads to a substantially stronger mass enhancement. Consequently, our DFT+DMFT simulations based on DSCC results demonstrate a higher level of agreement with the experimental data, which further validates the accuracy of the correlations parameters derived via DSCC calculations.

The renormalization of quasiparticles can be directly perceived from the spectral function presented in Fig. \ref{bands_specfuncs}(d) to (f).
Near the Fermi level, the momentum-resolved spectral function of SrIrO$_3$ shows minor deviations from the DFT band in Fig. \ref{bands_specfuncs}(a), indicating a relatively small mass enhancement of the Dirac quasiparticles near the Dirac points at M and A.
However, as can be seen from the spectral function of BaIrO$_3$, the dispersion at the Dirac points is strongly compressed, suggesting a significant mass enhancement for the Dirac quasiparticles. Additionally, the introduction of Coulomb correlation causes a notable change in the partial DOS near the Dirac points: the components of $J_{\mathrm{eff}}= 1/2$ orbital in SrIrO$_3$, BaIrO$_3$ and CaIrO$_3$ increase significantly, while the contributions of the other orbitals are significantly suppressed compared to the results of DFT. This indicates that these Dirac quasiparticles are almost entirely composed of the orbital with $J_{\mathrm{eff}}= 1/2$.

\section{topological surface states}

Dirac semimetals can generate observable topological surface states. For instance, on certain surfaces of Na$_3$Bi and Cd$_3$As$_2$, Fermi arcs connect two Dirac points and form closed loops, which are significantly different from the non-closed Fermi arcs in Weyl semimetals~\cite{Yang14,Wang12,Wang13}. To investigate the surface-state characteristics of SrIrO$_3$, BaIrO$_3$, and CaIrO$_3$, we constructed their tight-binding models using the Wannier90 software package~\cite{Pizzi20} and calculated the electronic dispersion relations of 60-layer thin films with (001) surfaces using the WannierTools software package~\cite{Wu17}. The results are shown in Figs. \ref{Slab}(b) to (d), where the gray and red lines correspond to the energy spectra of the bulk and surface states, respectively. For SrIrO$_3$ and BaIrO$_3$, the Brillouin zone of their (001) surfaces is shown in Fig. \ref{Slab} (a). On this surface, the Dirac points M and A (and their equivalent points along the $k_z$ direction) of the bulk state project onto the $\bar{S}$ and $\bar{\Gamma}$ points, respectively. Therefore, both the $\bar{S}$ and $\bar{\Gamma}$ points are the projection overlap points of two equivalent Dirac points. All points on the $\Gamma$-A path of the bulk state project onto the $\bar{\Gamma}$ point of the (001) surface. Thus, as can be seen from Figs. \ref{Slab}(b) and (c), the bulk state at the $\bar{\Gamma}$ point covers a relatively large energy range above and below the Fermi level (reflecting the energy dispersion of the bulk band along the $k_z$ direction, corresponding to the energy range from the Dirac point A to the $\Gamma$ point in Figs. \ref{bands_specfuncs}(a) and (b)), and completely covers the surface state within this range. Only around 0.2 eV above the Fermi level, the surface state is well separated from the bulk state. In contrast, the M-Y path projects onto the $\bar{S}$ point. Due to the smaller energy range of this segment, the energy range covered by the bulk state on the (001) surface is also smaller, thereby exposing some surface state dispersions that are not covered by the bulk state.

From Figs.\ref{Slab} (b) and (c), it can be observed that at the $\bar{S}$ and $\bar{\Gamma}$ points in the (001) surface Brillouin zone, the surface states intersect, forming closed Fermi arcs. This phenomenon can be intuitively demonstrated by the surface Fermi surface patterns of SrIrO$_3$ and BaIrO$_3$ shown in Figs.\ref{Slab}(e) to (g). Similar surface states and Fermi arcs have also been observed in other Dirac nodal line semimetals, such as CeSbTe and ZrSiS~\cite{Schoop18,Chen17}. Additionally, we found that the dispersion characteristics of these surface states are independent of the number of layers in the slab structure, indicating that these surface states are robust and topologically protected. For CaIrO$_3$, its Dirac nodal rings are small and are located near the S point on the bulk Brillouin zone boundary (see Fig.\ref{Slab}(h)). The energy dispersion of CaIrO$_3$ on the (001) surface is shown in Fig.\ref{Slab}(d), where multiple surface states appear in the projected area near the $\bar{S}$ point.

\section{conclusion}

Based on the Coulomb correlation parameters self-consistently calculated using the doubly screened Coulomb correction approach, this study employs a combined approach of density functional theory (DFT) and dynamical mean-field theory (DMFT) to systematically investigate the electronic structure and topological quantum states of MIrO$_3$ (M=Sr, Ba and Ca) compounds with nonsymmorphic symmetries. Through diagonalization analysis of the matrix of impurity levels for Ir-$5d$ orbitals, it is revealed that under the synergistic effect of strong spin-orbit coupling (SOC) and crystal fields in Ir-$5d$ states, orbital state reconstruction occurs, forming characteristic orbitals with an effective angular momentum of  $J_{\text{eff}}$ = 1/2. These orbital components constitute the dominant contribution to the energy bands near the Fermi level. DFT+DMFT calculations demonstrate that Coulomb correlation effects significantly enhance the weight proportion of  $J_{\text{eff}}$ = 1/2 states in the energy bands around the Fermi energy. Further investigations show that regulated by the nonsymmorphic symmetries of the MIrO$_3$ lattices,  $J_{\text{eff}}$ = 1/2 states form Dirac nodes near specific high-symmetry points in the Brillouin zone. Quasiparticle effective mass calculations based on DFT+DMFT indicate that these Dirac quasiparticles exhibit weakly correlated characteristics, with their dynamic behavior being in excellent agreement with experimental measurements. Additionally, through tight-binding model simulation of surface states, it is found that near high-symmetry points in the surface Brillouin zone, surface states are clearly decoupled from bulk states, forming closed surface Fermi arc structures. This work clarifies that in MIrO$_3$ systems with nonsymmorphic symmetries, the synergistic effects of symmetry protection, strong spin-orbit coupling, and electron correlations can induce unique topological states, providing a theoretical reference for the investigation of exotic correlated topological materials.

\acknowledgments
H. Li acknowledgements the supports from National Natural Science Foundation of China (No. 12364023), and Guangxi Natural Science Foundation (No. 2024GXNSFAA010273).


\begin{thebibliography}{99}


\bibitem{Lv15} B. Q. Lv, H. M. Weng, B. B. Fu, X. P. Wang, H. Miao, J.
Ma, P. Richard, X. C. Huang, L. X. Zhao, G. F. Chen, Z.
Fang, X. Dai, T. Qian, and H. Ding, \textit{Phys. Rev. X} \textbf{5}, 031013
(2015).

\bibitem{Wang19} Z. Wang, K. Luo, J. Zhao, and R. Yu, \textit{Phys. Rev. B} \textbf{100}, 205117
(2019).

\bibitem{Wang12} Zhijun Wang, Yan Sun, Xing-Qiu Chen, Cesare Franchini, Gang Xu, Hongming Weng, Xi Dai, and Zhong Fang,\textit{Phys. Rev. B} \textbf{85}, 195320 (2012).
%Dirac semimetal and topological phase transitions in A3Bi (A = Na, K, Rb)


\bibitem{Wang13} Zhijun Wang, Hongming Weng, Quansheng Wu, Xi Dai, and Zhong Fang, \textit{Phys. Rev. B} \textbf{88}, 125427 (2013).
%Three-dimensional Dirac semimetal and quantum transport in Cd3As2

\bibitem{Lai18} H.-H. Lai, S. E. Grefe, S. Paschen, and Q. Si, \textit{Proc. Natl. Acad.
Sci. USA} 115, \textbf{93} (2018).

\bibitem{Cao20} C. Cao, G.-X. Zhi, and J.-X. Zhu, \textit{Phys. Rev. Lett.} \textbf{124}, 166403
(2020).

\bibitem{Kofuji21} A. Kofuji, Y. Michishita, and R. Peters, \textit{Phys. Rev. B} \textbf{104},
085151 (2021).

\bibitem{Dzsaber21} S. Dzsaber, X. Yan, M. Taupin, G. Eguchi, A. Prokofiev, T.
Shiroka, P. Blaha, O. Rubel, S. E.Grefe,H.-H. Lai, Q. Si, and S.
Paschen, \textit{Proc. Natl. Acad. Sci. USA} \textbf{118}, e2013386118 (2021).

\bibitem{Ma23} Hao-Tian Ma, Xing Ming, Xiao-Jun Zheng, Jian-Feng Wen, Yue-Chao Wang, Yu Liu, and Huan Li, \textit{Phys. Rev. B} \textbf{107}, 075124 (2023).
%Node-line Dirac semimetal manipulated via Kondo mechanism in nonsymmorphic CePt2Si2

\bibitem{Kim08} B. J. Kim, Hosub Jin, S. J. Moon, J.-Y. Kim, B.-G. Park, C. S. Leem, Jaejun Yu, T.W. Noh, C. Kim, S.-J. Oh, J.-H. Park, V. Durairaj, G. Cao, and E. Rotenberg
\textit{Phys. Rev. Lett.} \textbf{101}, 076402 (2008).
%Novel Jeff=1/2 Mott State Induced by Relativistic Spin-Orbit Coupling in Sr2IrO4

\bibitem{Zhang13} Hongbin Zhang, Kristjan Haule, and David Vanderbilt, \textit{Phys. Rev. Lett.} \textbf{111}, 246402 (2013).
%Effective J=1/2 Insulating State in Ruddlesden-Popper Iridates: An LDA+DMFT Study

\bibitem{Fujiyama12} S. Fujiyama, H. Ohsumi, T. Komesu, J. Matsuno, B. J. Kim, M. Takata, T. Arima, and H. Takagi,
\textit{Phys. Rev. Lett.} \textbf{108}, 247212 (2012).
%Two-Dimensional Heisenberg Behavior of Jeff=2 Isospins in the ParamagneticState of the Spin-Orbital Mott Insulator Sr2IrO4

\bibitem{Samanta24} Subhasis Samanta, Dukgeun Hong, and Heung-Sik Kim,
\textit{Nanomaterials} \textbf{14}, 9 (2024).
%Electronic Structures of Kitaev Magnet Candidates RuCl3 and RuI3

\bibitem{Cao18} Gang Cao and Pedro Schlottmann,
\textit{Rep. Prog. Phys.} \textbf{81}, 042502 (2018).
%The challenge of spin¨Corbit-tuned ground states in iridates: a key issues review

\bibitem{Young15} S. M. Young and C. L. Kane, \textit{Phys. Rev. Lett.} \textbf{115}, 126803
(2015).

\bibitem{Fu17} B.-B. Fu, C.-J. Yi, T.-T. Zhang, M. Caputo, J.-Z. Ma, X.
Gao, B. Q. Lv, L.-Y. Kong, Y.-B. Huang, M. Shi et al.,
\textit{arXiv}:1712.00782.

\bibitem{Schoop15} L. M. Schoop, M. N. Ali, C. Stra{\ss}er, A. Topp, A. Varykhalov,
D. Marchenko, V. Duppel, S. S. P. Parkin, B. V. Lotsch, and
C. R. Ast, \textit{Nat. Commun.} \textbf{7}, 11696 (2015).

\bibitem{Wu22} H. Wu, A. M. Hallas, X. Cai, J. Huang, J. S. Oh, V. Loganathan,
A. Weiland, G. T. McCandless, J. Y. Chan, S.-K. Mo et al., \textit{npj
Quantum Mater.} \textbf{7}, 31 (2022).

\bibitem{Chen17} C. Chen, X. Xu, J. Jiang, S.-C. Wu, Y. P. Qi, L. X. Yang, M. X. Wang, Y. Sun, N. B. M. Schr\"{o}ter,
H. F. Yang, L. M. Schoop, Y. Y. Lv, J. Zhou, Y. B. Chen, S. H. Yao, M. H. Lu, Y. F. Chen, C. Felser,
B. H. Yan, Z. K. Liu, and Y. L. Chen, \textit{Phys. Rev. B} \textbf{95}, 125126 (2017).
%surface state of DNLs

\bibitem{Takayama18} T. Takayama, A. N. Yaresko, and H. Takagi, \textit{J. Phys.: Condens. Matter} \textbf{31}, 074001 (2018).
%Monoclinic SrIrO3 ¨C A Dirac semimetal produced by non-symmorphic symmetry and spin-orbit coupling

\bibitem{Hsu21} Yu-Te Hsu, Danil Prishchenko, Maarten Berben, Matija \v{C}ulo, Steffen Wiedmann, Emily C. Hunter, Paul Tinnemans, Tomohiro Takayama, Vladimir Mazurenko, Nigel E. Hussey, and Robin S. Perry,
\textit{npj Quantum Mater.} \textbf{6}, 92 (2021).
%Evidence for strong electron correlations in a nonsymmorphic Dirac semimetal

\bibitem{Tsuchiya07} Taku Tsuchiya and Jun Tsuchiya, \textit{Phys. Rev. B} \textbf{76}, 144119 (2007).
%Structure and elasticity of pnma CaIrO_ {3} and their pressure dependences: Ab initio calculations

\bibitem{Zeb12} M. Ahsan Zeb, and Hae-Young Kee, \textit{Phys. Rev. B} \textbf{86}, 085149 (2012).
%Interplay between spin-orbit coupling and Hubbard interaction in SrIrO3 and related Pbnm perovskite oxides

\bibitem{Haule10} K. Haule, C.-H. Yee, and K. Kim, \textit{Phys. Rev. B} \textbf{81}, 195107
(2010).

\bibitem{Blaha20} P. Blaha, K. Schwarz, F. Tran, R. Laskowski, G. K. H. Madsen,
and L. D. Marks, \textit{J. Chem. Phys.} \textbf{152}, 074101 (2020).

\bibitem{Arita12} R. Arita, J. Kune\v{s}, A.V. Kozhevnikov, A. G. Eguiluz, and M. Imada,
\textit{Phys. Rev. Lett.} \textbf{108}, 086403 (2012).
%Ab initio Studies on the Interplay between Spin-Orbit Interaction and Coulomb Correlation in Sr2IrO4 and Ba2IrO4

\bibitem{Tian24} Peng-Fei Tian, Hao-Tian Ma, Xing Ming, Xiao-Jun Zheng and Huan Li,
\textit{J. Phys.: Condens. Matter} \textbf{36}, 355602 (2024).
%Effective model and electroncorrelations in trilayer nickelatesuperconductor La4Ni3O10

\bibitem{Liu23} B.-L. Liu, Y.-C. Wang, Y. Liu, H.-F. Liu, and H.-F. Song, Doubly Screened Coulomb Correction Approach for Strongly Correlated Systems, \textit{J. Phys. Chem. Lett.} \textbf{14}, 8930 (2023).

\bibitem{Zhang17} Hongbin Zhang, Kristjan Haule, and David Vanderbilt, \textit{Phys. Rev. Lett.} \textbf{118}, 026404 (2017).
%Metal-Insulator Transition and Topological Properties of Pyrochlore Iridates

\bibitem{Imai22} Yoshinori Imai, Kazuhiro Nawa, Yasuhiro Shimizu, Wakana Yamada, Hideyuki Fujihara, Takuya Aoyama, Ryotaro Takahashi, Daisuke Okuyama, and Takamasa Ohashi et al.,
\textit{Phys. Rev. B} \textbf{105}, L041112 (2022).
%Zigzag magnetic order in the Kitaev spin-liquid candidate material with a honeycomb lattice

\bibitem{Kaib22} David A. S. Kaib, Kira Riedl, Aleksandar Razpopov, Ying Li, Steffen Backes, Igor I. Mazin, and Roser Valent\'{\i}
\textit{npj Quantum Materials} \textbf{7}, 75 (2022).
%Electronic and magnetic properties of the RuX3 (X = Cl, Br, I) family: two siblings¡ªand a cousin?

\bibitem{Zhang22} Yang Zhang, Ling-Fang Lin, Adriana Moreo, and Elbio Dagotto,
\textit{Phys. Rev. B} \textbf{105},085107 (2022).
%Theoretical study of the crystal and electronic properties of a-RuI3

\bibitem{Nawa21} Kazuhiro Nawa, Yoshinori Imai, Youhei Yamaji, Hideyuki Fujihara,
Wakana Yamada, Ryotaro Takahashi, Takumi Hiraoka, Masato Hagihala,
Shuki Torii, Takuya Aoyama, Takamasa Ohashi, Yasuhiro Shimizu,
Hirotada Gotou, Masayuki Itoh, Kenya Ohgushi, and Taku J. Sato,
\textit{Journal of the Physical Society of Japan} \textbf{90}, 123703 (2021).
%Strongly Electron-Correlated Semimetal RuI3 with a Layered Honeycomb Structure

\bibitem{Kim15} Heung-Sik Kim, Vijay Shankar V., Andrei Catuneanu, and Hae-Young Kee, \textit{Phys. Rev. B} \textbf{91}, 241110(R) (2015).
%Kitaev magnetism in honeycomb RuCl3 with intermediate spin-orbit coupling

\bibitem{Yang14} Bohm-Jung Yang and Naoto Nagaosa, \textit{Nat. Commun.} \textbf{5}, 4898 (2014).
%Classification of stable three-dimensional Dirac semimetals with nontrivial topology

\bibitem{Pizzi20} G. Pizzi, V. Vitale, R. Arita, S. Bl\"{u}gel, F. Freimuth, G. G\'{e}ranton, M. Gibertini, D. Gresch, C. Johnson, T. Koretsune, J. Iba\~{n}ez-Azpiroz, H. Lee, J. M. Lihm, D. Marchand, A. Marrazzo, Y. Mokrousov, J.I. Mustafa, Y. Nohara, Y. Nomura, L. Paulatto, S. Ponc\'{e}, T. Ponweiser, J. Qiao, F. Th\"{o}le, S.S. Tsirkin, M. Wierzbowska, N. Marzari, D. Vanderbilt, I. Souza, A. A. Mostofi, J. R. Yates, \textit{J. Phys. Cond. Matt.} \textbf{32}, 165902 (2020).
%Wannier90 as a community code: new features and applications,

\bibitem{Wu17} QuanSheng Wu, ShengNan Zhang, Hai-Feng Song, Matthias Troyer, Alexey A Soluyanov, \textit{Computer Physics Communications} \textbf{224}, 405 (2017).

\bibitem{Schoop18} Leslie M. Schoop, Andreas Topp, Judith Lippmann, Fabio Orlandi, Lukas M\"{u}chler,
Maia G. Vergniory, Yan Sun, Andreas W. Rost, Viola Duppel, Maxim Krivenkov,
Shweta Sheoran, Pascal Manuel, Andrei Varykhalov, Binghai Yan, Reinhard K. Kremer,
Christian R. Ast, Bettina V. Lotsch, \textit{Sci. Adv.} \textbf{4} eaar2317 (2018).
%surface state of DNLs








\end{thebibliography}
\end{document}